\documentclass[aps,prl,twocolumn]{revtex4}
\newcommand{\bea}{\begin{eqnarray}}
\newcommand{\eea}{\end{eqnarray}}
\newcommand{\pslash}{\not\hspace{-0.7mm}p}
\newcommand{\kslash}{\not\hspace{-0.7mm}k}
\usepackage{amsmath}
\usepackage{graphicx}
\usepackage{bm}
\DeclareGraphicsExtensions{.ps}
\begin{document}

\title{Alberg and Miller Reply}

\author{Mary Alberg$^{\, 1,2}$, Gerald A. Miller${\, ^2}$}
\affiliation{
   $^1$ Department of Physics, % Box 8202,
	Seattle University, Seattle, Washington 98122-1090, USA	\\
   $^2$ Department of Physics, University of Washington, Seattle, Washington 98195-1560, USA}

\preprint{NT@UW-13-16}

\maketitle

%%%%%%%%%%%%%%%%%%%%%%%%%%%%%%%%%%%%%%%%%%%%%%%%%%%%%%%%%%%%%%%%%%%%%%%%%
In their comment Ji {\it et al.}~\cite{ji}  correctly state that we ``obtain the same result as given by the PV theory" for the self-energy of the nucleon.
 This means that our principal conclusion, that ``The pion mass $\mu$  dependence of  $\Sigma_\pi$ is  consistent with chiral perturbation theory results for small values of $\mu$ and  is  also   linearly dependent on  $\mu$ for larger values,  in accord with the results of  lattice
QCD calculations", is correct. A consequence of the latter result is that the chiral  limit is not yet numerically relevant for
 current QCD lattice  calculations.

Ji {\it et al.} conclude by stating ``The pseudoscalar coupling therefore cannot in general be used if one  wishes to ensure consistency with the chiral properties of QCD which are respected by the pseudovector $\pi N$ coupling". This conclusion is obviously correct, but no claim
of general equivalence between PS and PV theory was made in our paper~\cite{Alberg12}. 

The conclusion of Ji {\it et al.} is driven by their opening statement  that  we argued:
``form factors suppress the off-shell contact interactions". No such argument is given in our paper. They further state
that our results are obtained by neglecting end-point singularities at $k^+=0$. This is not the case. Our
results are obtained by putting the intermediate nucleon on its mass shell. With this procedure the self energy obtained with PS and PV theories is indeed the same, as is expected from the well-known equivalence theorem. The use of on-mass shell intermediate nucleons allows the  use of experimentally measured $\pi N$ form factors, which greatly reduces the uncertainty of the result.

Our pion-nucleon effective theory uses PV interactions without approximations, as we demonstrate.  The key point is the use of the identity~\cite{Yan}  for the nucleon propagator:
\bea {1\over \pslash-\kslash +M}= {\sum_s u(p-k,s)\bar{u}(p-k,s)\over (p-k)^2 -M^2}+{\gamma^+\over 2(p-k)^+},\eea
where $u(p-k,s)$ is an on-shell Dirac light-front spinor, $p$ is the initial momentum of the nucleon and $k$ is the momentum of the virtual pion. Using  the first term  gives the same result for PS  or PV interactions, and  we show that  the second term vanishes. This term involves the factor $\gamma^5\kslash\gamma^+ \gamma^5\kslash/(2(p^+-k^+)$ which is 
evaluated as $1/(2M)k_\perp^2p^+/(p^+-k^+)$ in the infinite momentum frame
$(p^+\to\infty)$. The integral over $k_\perp^2$ involves a well-behaved integrand if our  our form factors are included, and one finds  the result:
\bea
\Sigma\sim\int dk^+dk^- {p^+\over p^+-k^+}{\cal F}(k^+k^-)
.\eea where ${\cal F}(k^+k^-)$ is a well-behaved function. The integral can be performed by using $y=k^+k^-$ and $k^+$ as the integration variables.
The result is that
\bea
\Sigma\sim\int {dk^+\over k^+} {p^+\over p^+-k^+}\int dy {\cal F}(y)=0. 
.\eea The principal values integration over $k^+$ gives 0. This is not happenstance.
The Ji {\it et al.}  finding  that removing end-point singularities from the PS  calculation yields the PV result is accidental.

The only possible change in our numerical results arising from the effects discussed by Ji {\it et al.} is on the pion distribution  $f_\pi^N$.
However, the  work of Refs.~\cite{Thomas00,Detmold01} shows that the effect is to  increase the value of a term that has a very  small contribution  (when realistic values of the  pion mass are used)
by a factor of 4/3. The size of this effect is much, much  less than the uncertainty introduced by the use of  $\pi N$ form factors that are not constrained by measurements. 

To summarize, nothing in the Comment by Ji {\it et al.} changes the conclusions or numerical results of \cite{Alberg12} in any substantive way.

 \vspace*{0.5cm}

This work was partially supported by the DOE contract No. DE-FG02-97ER-41014 and NSF Grant
No. 0855656.

% \vspace*{-0.5cm}

%%%%%%%%%%%%%%%%%%%%%%%%%%%%%%%%%%%%%%%%%%%%%%%%%%%%%%%%%%%%%%%%%%%%%%%%%


\begin{thebibliography}{99}

% \vspace*{-0.5cm}

\bibitem{ji} C-R Ji, W. Melnitchouk, A.W. Thomas,
previous comment

\bibitem{Alberg12}
M.~Alberg and G.~A.~Miller,
Phys.\ Rev.\ Lett.\  {\bf 108}, 172001 (2012).


\bibitem{Yan} 
  S.~-J.~Chang and T.~-M.~Yan,
  %``Quantum field theories in the infinite momentum frame. 2. Scattering matrices of scalar and Dirac fields,''
  Phys.\ Rev.\ D {\bf 7}, 1147 (1973).
  %%CITATION = PHRVA,D7,1147;%%
\bibitem{Thomas00}
% A.~W.~Thomas {\it et al.}, % W.~Melnitchouk and F.~M.~Steffens,
A.~W.~Thomas, W.~Melnitchouk and F.~M.~Steffens,
Phys.\ Rev.\ Lett.\  {\bf 85}, 2892 (2000).

\bibitem{Detmold01}
% W.~Detmold {\it et al.},
W.~Melnitchouk, J.~W.~Negele, D.~B.~Renner and A.~W.~Thomas,
Phys.\ Rev.\ Lett.\  {\bf 87}, 172001 (2001).


\end{thebibliography}
\end{document}